\documentclass{moriond}

\def\be{\begin{equation}}
\def\ee{\end{equation}}
\def\bea{\begin{eqnarray}}
\def\eea{\end{eqnarray}}

\newcommand{\Photo}{}

\begin{document}
\vspace*{4cm}
\title{Direct Dark Matter Search with the CRESST-II Experiment}
\author{J. Schieck \\ for the CRESST collaboration}
\address{Institute of High Energy Physics, Nikolsdorfer Gasse 18, 1050 Wien, Austria and\\
University of Technology, Institute of Atomic and Subatomic Physics, Stadionallee 2, 1020 Wien, Austria}

\maketitle\abstracts{
The quest for the particle nature of dark matter is one of the big open questions of modern physics. 
The CRESST-II experiment, located at the Gran Sasso laboratory in Italy, is optimised for the detection 
of the elastic scattering of dark matter particles with ordinary matter. We present the result 
obtained with an improved detector setup with increased radiopurity and enhanced background rejection. 
The limit obtained in the so-called low mass region between one and three GeV/$c^{2}$ is at the present 
among the best limits obtained for direct dark matter experiments. In addition we give an outlook of the
 future potential for direct dark matter detection using further improved CRESST CaWO$_{4}$ cryogenic detectors.
}

\section{Introduction}
Understanding the origin of dark matter is one of the scientific key questions at the beginning of the 21$^{\mathrm{st}}$ century. The
existence of dark matter at galactic and cosmological scales is undisputed. Several uncorrelated measurements indicate that
about $25\%$ of the overall energy-matter density of the universe consists of Dark Matter~\cite{DelPopolo:2013qba}. The underlying
character of dark matter is still not understood. Assigning particle character to dark matter is one of the most promising approaches
for solving this open topic. However, none of the known particles of the Standard Model of particle physics 
can act as a candidate for Dark Matter. 
The new particle cannot be charged or baryonic, it must be cold, traveling much more slow than the speed of light, and its decaytime
must be below the age of the universe. The abundance of the dark matter particles should match the dark matter relic density.
The interaction of a hypothetic dark matter particle with known Standard Model particles is 
unknown. All dark matter observations are based on gravity, however, a weak interaction between the unknown dark matter particle and 
known particles is expected. \\
One of the best studied particle candidates for dark matter is the so-called WIMP, a weakly interacting massive particle. WIMPs are 
thermally produced during the evolution of the universe. A WIMP dark matter candidate with a mass at the electroweak mass scale 
($\mathcal{O} (100 \mathrm{GeV}/c^{2})$)and a cross section similar to the electroweak scale ($\mathcal{O} (\mathrm{pb})$) could match 
the observed relic dark matter density. However, thermally produced WIMPs with a mass below 5-10 GeV/$c^{2}$ would lead to an over-closure 
of the universe and are therefore excluded. The best exclusion limits for WIMPs are obtained with direct dark matter detection experiments 
based on liquid noble gases.    \\
Asymmetric dark matter models predict particle candidates below the mass regime expected 
from standard WIMP based models~\cite{Zurek:2013wia}. 
The underlying model links two big unsolved riddles of modern physics: dark matter and the baryon asymmetry. A baryon violating process 
leads to a small imbalance between matter and antimatter and during the evolution of the universe the remaining antimatter completely 
annihilates with the matter, leading to the observed matter dominated environment. 
The observed dark matter and baryon density in the universe is of similar magnitude, motivating the assumption that the process leading
to the observed matter-anti-matter imbalance also controls the dark matter density. Asymmetric dark matter models predict anti-dark 
Matter and matter being initially equally produced. The same process leading to the observed baryon asymmetry is expected to 
produce a small excess of dark matter over anti-dark Matter, succeeded by the annihilation of the anti-dark Matter component. Following
this idea, we expect that the mass of the hypothetical dark matter particle is about five times larger than the typical mass scale of 
ordinary matter, which is dominated by the proton mass. Asymmetric dark matter models predict Dark Matter
candidates around 5 GeV/$c^{2}$,  below the mass scale expected for WIMPs. \\
This article summarises the results previously published in~\cite{Angloher:2014myn}.
\subsection{Search for low mass dark matter candidates}
Most direct dark matter detection experiments aim for the identification of dark matter candidates by elastic scatters 
with a target nucleus. The nuclear recoil 
deposits energy in the experiment and the amount of the energy is a measure of the mass of the dark matter candidate. The 
expected differential recoil energy spectrum $d\mathrm{R}/d\mathrm{E_{R}}$, with R being the event rate and $\mathrm{E_{R}}$
the deposited recoil energy, falls exponentially. The $d\mathrm{R}/d\mathrm{E_{R}}$ spectrum falls steeper for low mass dark matter candidates compared to heavy candidates. For an increased sensitivity for low mass dark matter particles a experiment needs to 
extend the recoil energy detection threshold towards lower values.
\subsection{Results from the previous CRESST-II Data taking Period}
\label{Result2011}
In 2011 the CRESST collaboration reported an excess of events over the
expected number of background events~\cite{Angloher:2011uu}. One
background source originated from polonium decays, $^{210}$Po$\to^{206}$Pb + $\alpha$. The lead nuclei
from the polonium decay can deposit energy in the target crystal with the generated signal being similar to 
the signal of a dark matter candidate. Interpreting the surplus of events in the region of interest as dark matter elastic
scattering resulted in two distinguished solutions, M1 and M2, in the mass versus cross section plane, with a significance 
of more than 4 $\sigma$. The analysis energy threshold for nuclear recoils was above 10 keV, leading to a steep decrease in 
sensitivity for dark matter candidates below 10 GeV.

\section{The CRESST-II Experiment}
Two main facts drive the design of a direct detection dark matter experiment: Dark matter particle - nucleus 
elastic scattering leads to a nuclear recoil 
with an energy deposition of a few keV at most. Second, dark matter candidates are expected to interact only weakly with the nucleus, 
 resulting in an extremely low interaction rate. For this reason a successful dark matter experiment needs to be sensitive to  very 
low energy depositions of a few keV or less, and an effective background rejection or suppression is needed for a good 
signal to background ratio. \\
To suppress background events originating from cosmic rays, the CRESST-II experiment is located at the Gran Sasso underground 
laboratory in Italy. In addition, the experiment contains active and passive background shielding to repress background events.
The result presented here is based on data taken in the year 2013.

\subsection{Cryogenic Detectors}

\begin{figure}[h]
\begin{minipage}{0.5\linewidth}
\centerline{\includegraphics[width=0.90\linewidth]{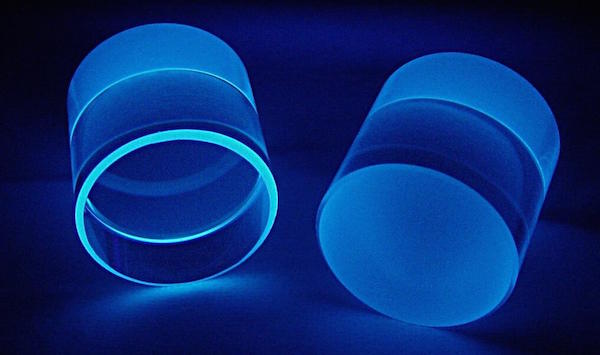}}
\end{minipage}
\hfill
\begin{minipage}{0.5\linewidth}
\centerline{\includegraphics[width=0.90\linewidth]{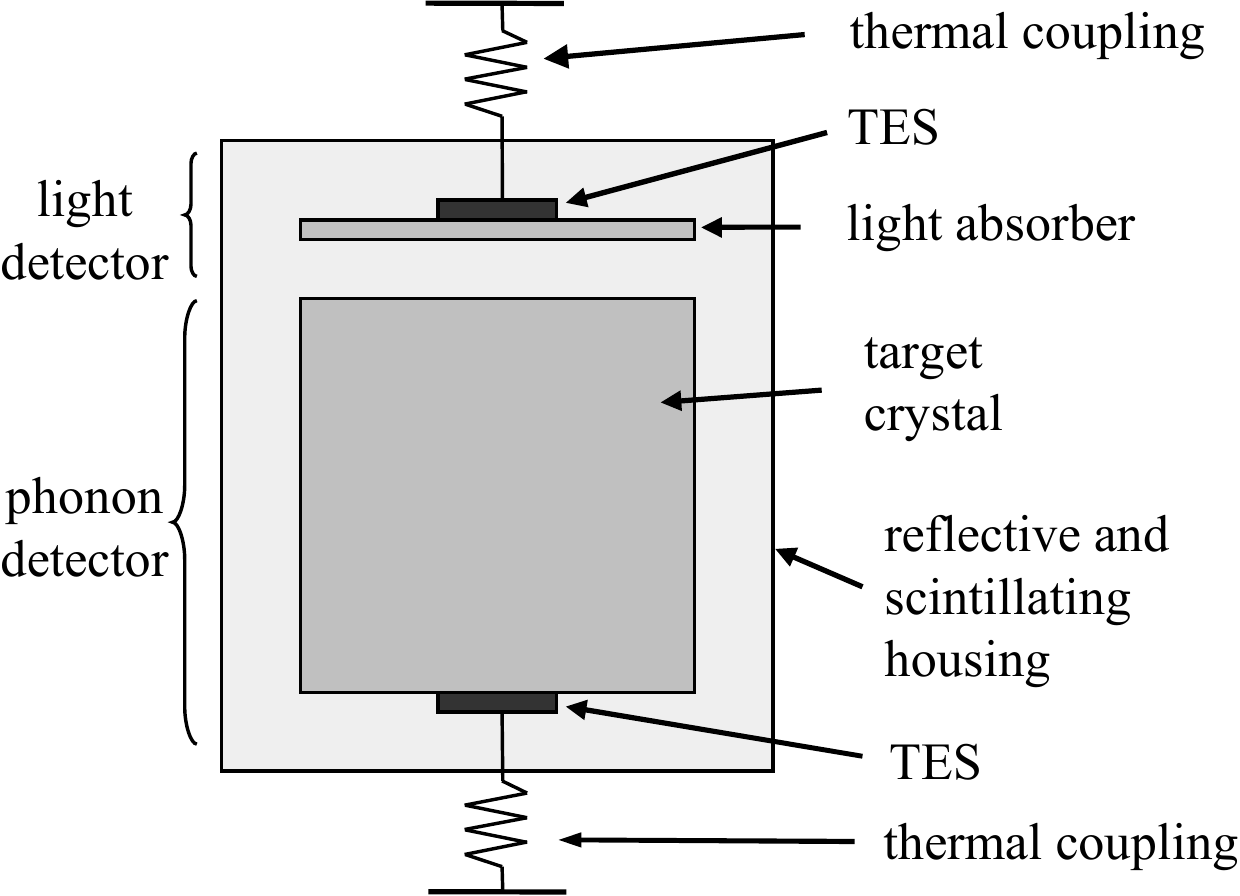}}
\end{minipage}
\hfill
\caption[]{Left: Light emission from two scintillating CaWO$_{4}$-crystals. Right: Sketch of a CRESST detector modul
with the target crystal being read out by a transition edge superconductor (TES) to detect the phonons and a light absorber
combined with another TES to measure the scintillation light. The whole crystal is thermally coupled 
to a heat bath and operated within a reflective scintillating housing~\cite{Angloher:2011uu}.}
\label{fig:module}
\end{figure}

\begin{figure}[h]
\begin{center}
\begin{minipage}{0.75\linewidth}
\centerline{\includegraphics[width=.85\linewidth]{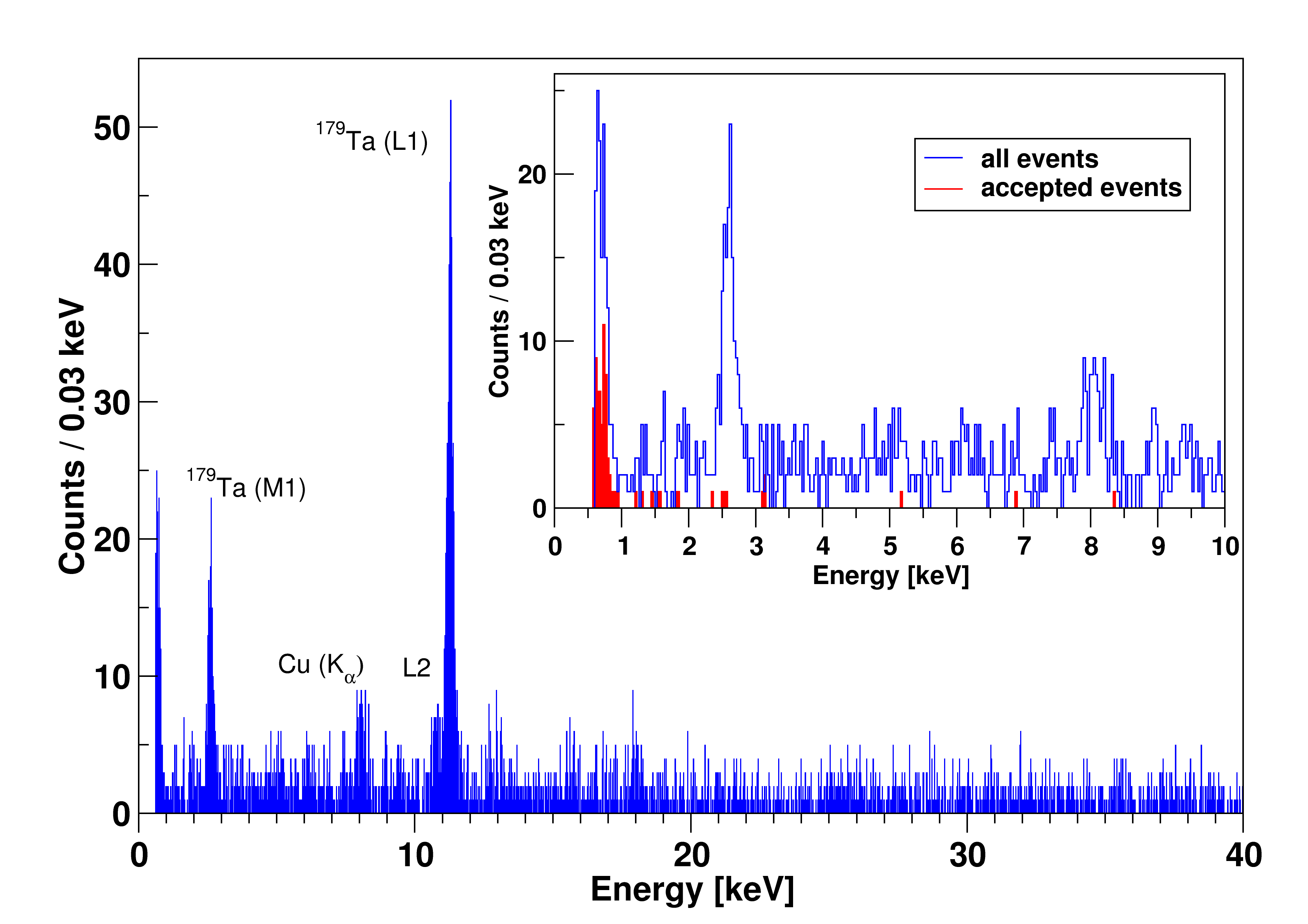}}
\end{minipage}
\caption[]{The energy distribution of all selected events. The indicated lines mainly originate from cosmogenic activation. The
insert is a zoom into the energy region below 10 keV energy with blue being all events and red indicating 
the events in the region of interest~\cite{Angloher:2014myn}.}
\label{fig:radiopurity}
\end{center}
\end{figure}

In CRESST-II the target material for the dark matter particle nucleus elastic scattering is CaWO$_{4}$, a scintillating crystal 
operated at a temperature of about 10 mK. This very low operating temperature results in a small heat capacity. A 
small energy deposition of about few keV returns a measurable temperature change of several $\mu$K. 
The crystal is thermally coupled to a heat bath. This temperature increase, emerging as phonons,
is detected in a bolometric mode by superconducting thermometers being operated at their phase transition 
between the normal  and the superconducting state. Almost the
complete deposited energy is carried away by the phonons allowing for an unbiased measurement of the full deposited energy.
A small fraction of the energy is transformed into scintillation light. The scintillation light is detected by a light detector made 
of silicon-on-saphire,  read out again by a superconductor operated at its phase transition. An impression of the scintillation light 
being emitted by a CaWO$_{4}$-crystal is visualised in Fig.~\ref{fig:module} (left).
The light emitted strongly depends on the type of radiation and the type of recoiling target. 
The so-called light yield, the energy deposited in the light channel normalised to the energy in the phonon channel, differs
for an electron or gamma  compared to nuclear recoils. The light yield of an electron or gamma 
is calibrated to one at 122 keV and a nuclear recoil returns a significantly reduced value of almost zero. A cut on the light 
yield therefore allows an efficient discrimination of background events originating from electrons or gammas. 
For an improved background rejection the crystal is located in a scintillating housing. 
Additional light from the alpha impinging on the housing veto events from surface alpha decays 
originated from the inner surfaces of the module.
As discussed above (see section~\ref{Result2011})
alpha-decays from polonium close to the crystal surface could mimic a signal event. 
During the previous data taking period the coverage was not hermetic and due to the missing 
scintillating light some $^{210}$Po related decays may not have been identified as background events. For the 
data taking period described here the module design was adjusted, leading 
to a complete coverage. In addition the mounting was performed in a radon free environment leading to
a decreased $^{210}$Po contamination of the surfaces.  A sketch of the module design is shown in~ Fig.\ref{fig:module} (right). \\
The CaWO$_{4}$ crystal used for data taking is produced at the Technische Universit\"at M\"unchen. Special care is taken 
in selecting radiopure material and in the various production steps to achieve a  very high intrinsic radiopurity. As a result the 
electron / gamma background in the region of interest could be improved up to a factor of ten with respect to commercially available crystals. 
The low-energy 
spectrum of a single module (TUM-40) with an exposure of 29.4 kg days is shown in Fig.~\ref{fig:radiopurity}.
The threshold energy for a $50\%$ trigger efficiency is determined to be about 600 eV, with a resolution of about 100 eV.
Various quality cuts are applied to the recorded events. Events are required to be collected during stable detector operation
and are not to be in  coincidence with cosmic muon events or with an event in any other detector module. 
The fit to determine the energy from the measured
signal pulse needs to fulfil certain quality criteria and the signal shape must be consistent with the energy being deposited in
the crystal and not in any other part of the detector module.
These cuts lead to an overall efficiency just above $70\%$ for energies above 3 keV.

\section{Results}
\begin{figure}[h]
\begin{center}
\begin{minipage}{0.75\linewidth}
\centerline{\includegraphics[width=.90\linewidth]{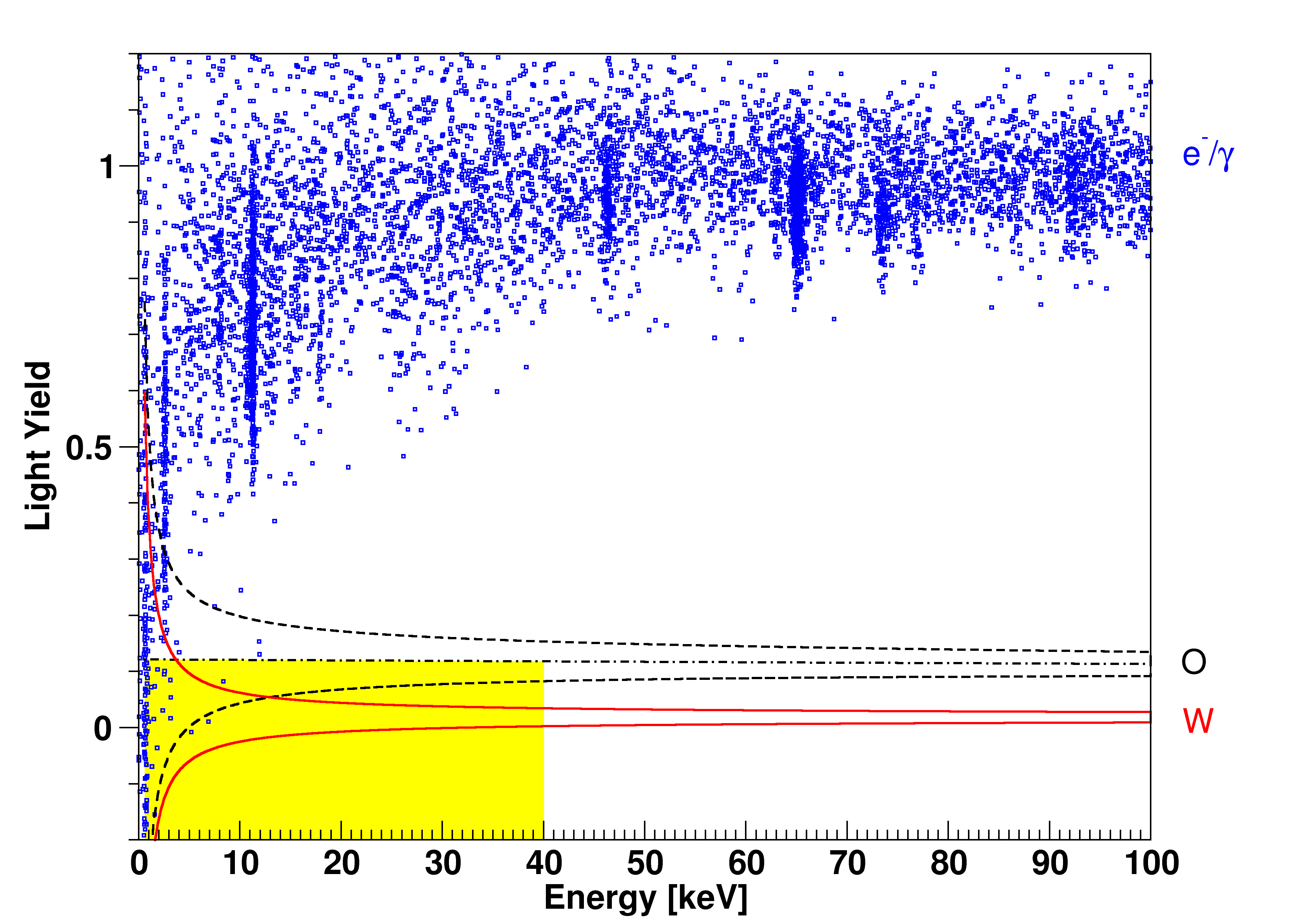}}
\end{minipage}
\caption[]{Light yield versus energy for events after the selection being applied. The
dashed grey line indicates the expected light yield for dark matter particle oxygen nucleus elastic scattering events (oxygen band) 
and the red line indicates the light yield for  dark matter particle tungsten nucleus scattering events (tungsten band). 
The region of interest is 
defined for nuclear recoils between 0.6 keV and 40 keV and  a light yield below the center of the oxygen band (yellow box)
~\cite{Angloher:2014myn}.}
\label{fig:eventsLightYield}
\end{center}
\end{figure}

\begin{figure}[h]
\begin{center}
\begin{minipage}{0.75\linewidth}
\centerline{\includegraphics[width=.900\linewidth]{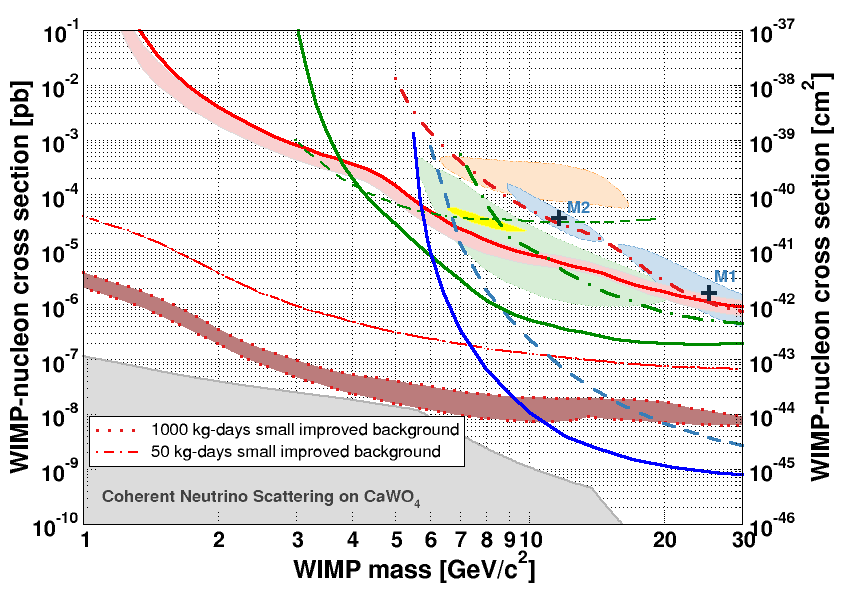}}
\end{minipage}
\caption[]{The exclusion limit for the dark matter nucleon cross section as a function
of the dark matter mass is shown here. The red line represents the limit obtained 
with data taken taken by the CRESST-II experiment in 2013. The dotted and dash-dotted lines indicate the
limits expected for future data taking using an improved CRESST-II detector module design and 
reduced intrinsic radioactive background. The other lines correspond to limits obtained
with other dark matter experiments, for a complete description of the other results 
see~\cite{Angloher:2014myn,Angloher:2015eza}.}
\label{fig:result}
\end{center}
\end{figure}

The distribution of events after the event selection criteria being applied is shown in Fig.~\ref{fig:eventsLightYield}. 
Background  electron / gamma events with a light yield being consistent with one are clearly visible. 
Towards smaller  energies the emitted scintillating light output is reduced and  
the noise of the light detector dominates the width of the light yield distribution. 
At  energies below about 10 keV electron / gamma events leak into the signal region, representing the
only background contribution.
The number of events observed in the region of interest (Fig.~\ref{fig:eventsLightYield}) is consistent
with the expected number of events originating from a leakage of electron / gamma events. 
All events are considered as  signal events and a limit of dark matter particle nucleon cross-section is set.
The spectrum of expected signal events is estimated using standard 
astrophysical assumptions for the dark matter distribution within the vicinity of the earth.
The upper limit for the cross section as a function of the dark matter particle mass
is calculated with Yellin's optimum interval method~\cite{Yellin:2002xd}. The exclusion limit for masses
between one and 30 GeV/$c^{2}$ is summarised in Fig.~\ref{fig:result}. The result previously obtained and 
if being interpreted as a possible WIMP signal~\cite{Angloher:2011uu}
indicated with M2 in Fig.~\ref{fig:eventsLightYield} is excluded and the point M1 is disfavoured.
\section{Outlook}
The limit reached by the CRESST-II experiment is among the best limits obtained for
direct detection of dark matter particles in the mass region between one and three GeV/$c^{2}$.
Further improvements in the module design will lead to a
reduced detection threshold, additionally improving the limit for
low mass dark matter particles~\cite{Angloher:2015eza}. Together with a 
reduction of the intrinsic radioactive contamination the sensitivity almost reaches the level of 
coherent solar neutrino nucleus scattering on CaWO$_{4}$. The expected sensitivity is summarised in Fig.~\ref{fig:result}.
\section{Historical Interlude}
\begin{figure}[h]
\begin{center}
\begin{minipage}{0.75\linewidth}
\centerline{\includegraphics[width=.900\linewidth]{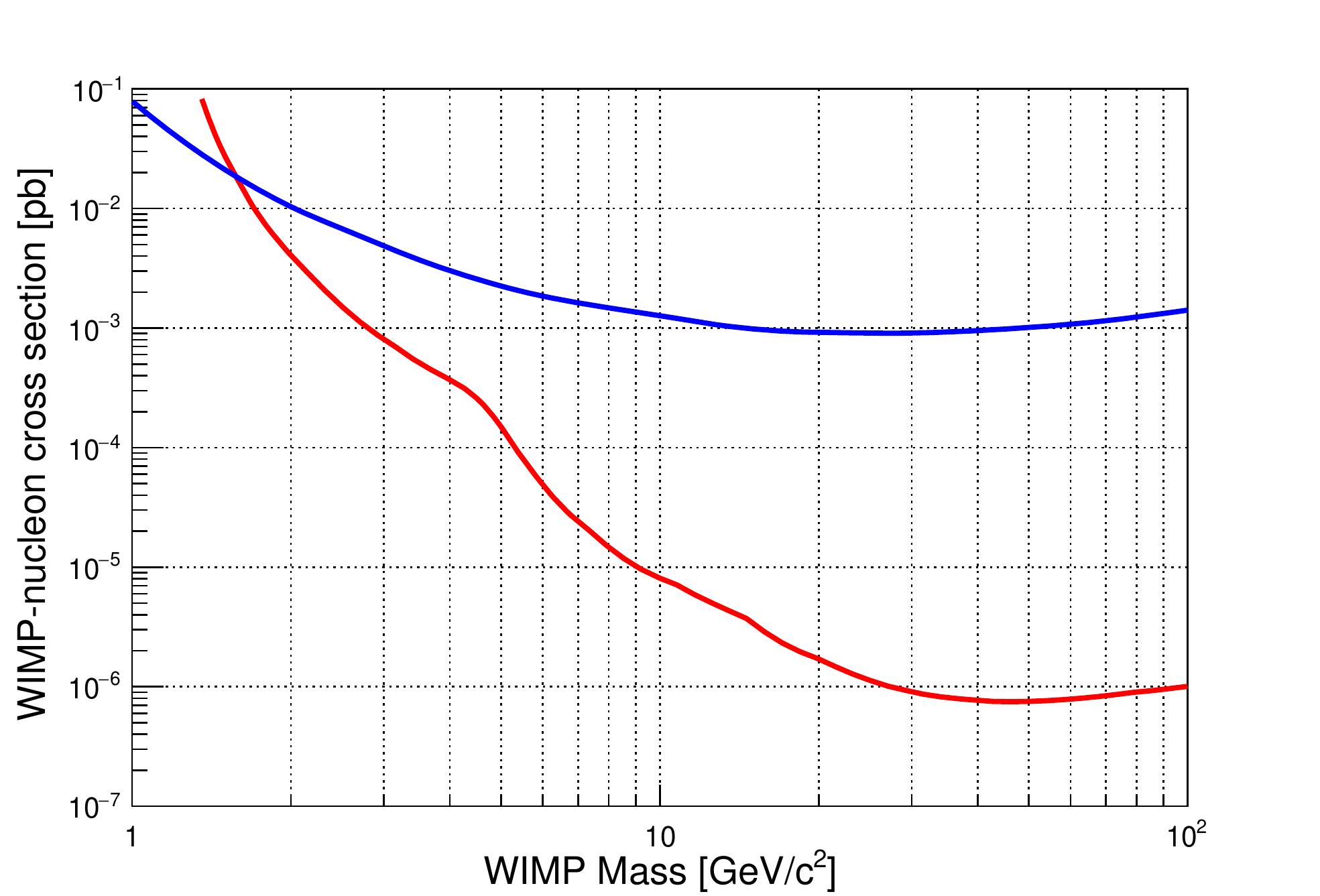}}
\end{minipage}
\caption[]{Limits for low mass dark matter particles 
obtained with the CRESST (blue)~\cite{Angloher:2002in} and the CRESST-II experiment (red)~\cite{Angloher:2014myn}.}
\label{fig:historical}
\end{center}
\end{figure}

The achieved sensitivity for low mass dark matter particle candidates is mainly due to the 
very low detection threshold and the high intrinsic radiopurity. 
Limits obtained with the predecessor experiment, the CRESST experiment~\cite{Angloher:2002in}, 
achieved a similar sensitivity in the low mass region (see Fig.~\ref{fig:historical}). The CRESST experiment
also used a cryogenic setup, however with Al$_{2}$O$_{3}$-crystals as target material. Al$_{2}$O$_{3}$ is not scintillating
and therefore the background discrimination is hindered, leading to a reduced sensitivity in the higher mass region.
With the current technology CRESST-II has the foremost limit in the mass region of low-mass dark matter particles, 
while still reaching a reasonable sensitivity for medium-mass dark matter particles. Increasing 
the exposure would allow a further improvement in sensitivity by up to four orders of magnitude in the medium- and
high-mass region.
\section{Summary}
CRESST-II is a cryogenic experiment searching for a dark matter particle elastically scattering from a nucleus 
of a CaWO$_{4}$ crystal. A new data taking campaign using an improved module design and increased 
radiopurity could not confirm the excess above background events previously observed by CRESST-II. The significantly 
reduced detection
threshold is about 600 eV, leading to high sensitivity for the detection of low-mass dark matter particles. Currently 
CRESST-II is among the direct dark matter detection experiments with the best limit in the low mass region. 
Future upgrades of the experiment
are expected to increase the sensitivity in this mass region by three to four orders of magnitude.

\end{document}